\documentclass{article}
\usepackage{spconf}
\usepackage{amsmath}
\usepackage{graphicx}  
\usepackage{amssymb}
\usepackage{multirow}
\usepackage{threeparttable}
\usepackage{cite}
\usepackage{booktabs}
\usepackage{makecell}
\usepackage{diagbox}
\usepackage{color}

\def\Vec#1{\textit{\boldmath $#1$}}
\def\Argmax{\mathop{\rm argmax}}

\def\red#1{#1}

\title{Zero- and Few-shot Sound Event Localization and Detection}
\name{Kazuki Shimada$^{\star\dagger}$, Kengo Uchida$^{\star}$, Yuichiro Koyama$^{\ddag}$, Takashi Shibuya$^{\star}$,}
\secondlinename{Shusuke Takahashi$^{\ddag}$, Yuki Mitsufuji$^{\star\ddag}$, Tatsuya Kawahara$^{\dagger}$}

\address{$^{\star}$ Sony \red{AI} \hspace{0.5cm} $^{\dagger}$ Kyoto University \hspace{0.5cm} $^{\ddag}$ Sony Group Corporation}

\begin{document}
\ninept

\maketitle

\begin{abstract}

Sound event localization and detection~(SELD) systems estimate direction-of-arrival~(DOA) and temporal activation for sets of target classes.
Neural network~(NN)-based SELD systems have performed well in various sets of target classes, but they only output the DOA and temporal activation of preset classes trained before inference.
To customize target classes after training, we tackle zero- and few-shot SELD tasks, in which we set new classes with a text sample or a few audio samples.
While zero-shot sound classification tasks are achievable by embedding from contrastive language-audio pretraining~(CLAP), zero-shot SELD tasks require assigning an activity and a DOA to each embedding, especially in overlapping cases.
To tackle the assignment problem in overlapping cases, we propose an embed-ACCDOA model, which is trained to output track-wise CLAP embedding and corresponding activity-coupled Cartesian direction-of-arrival~(ACCDOA).
In our experimental evaluations on zero- and few-shot SELD tasks, the embed-ACCDOA model showed better location-dependent scores than a straightforward combination of the CLAP audio encoder and a DOA estimation model.
Moreover, the proposed combination of the embed-ACCDOA model and CLAP audio encoder with zero- or few-shot samples performed comparably to an official baseline system trained with complete train data in an evaluation dataset.

\end{abstract}

\begin{keywords}
Sound event localization and detection (SELD), contrastive language-audio pretraining (CLAP)
\end{keywords}

\section{Introduction}
\label{sec:intro}

Given multichannel audio signals, a sound event localization and detection~(SELD) system simultaneously estimates the direction-of-arrival~(DOA) and temporal activation of target classes.
SELD plays an essential role in many applications, such as surveillance~\cite{crocco2016audio}, bio-diversity monitoring~\cite{chu2009environmental}, and smart devices~\cite{yalta2017sound,sun2021emergency}.
Each application has its own set of target sound event classes.
For example, a surveillance SELD system is required to detect and localize screams, gunshots, or glass breaking, whereas a smart home SELD system needs to detect and localize speech, footsteps, or dog barking.

Recent neural network~(NN)-based SELD systems usually set the target classes around ten sound events~\cite{politis2021dataset,politis2022starss22,adavanne2018sound,shimada2021accdoa,cao2021improved,slizovskaia2022locate,hu2023meta,kim2023ad,wang2023loss,niu2023experimental}.
After training with annotated multichannel audio data, the NN-based systems detect and localize target sound events.
There are two main approaches to associate a detection result with its DOA: class-wise and track-wise.
An example of the class-wise approach is SELDnet~\cite{adavanne2018sound}, which outputs an activity and a DOA of each target class.
An activity-coupled Cartesian DOA~(ACCDOA) vector assigns the activity to the length of a Cartesian DOA vector~\cite{shimada2021accdoa}, and the ACCDOA vector enables us to unify the activity and DOA branches into an ACCDOA branch.
Event independent network v2~(EINV2) is a track-wise method, in which each track estimates an event's class and the corresponding location~\cite{cao2021improved}.
These SELD systems have shown a reasonable performance in both simulated and real environments~\cite{politis2021dataset,politis2022starss22,adavanne2018sound,shimada2021accdoa,cao2021improved,slizovskaia2022locate,hu2023meta,kim2023ad,wang2023loss,niu2023experimental}.

\begin{figure}[t]
    \centering
    \centerline{\includegraphics[width=0.88\linewidth]{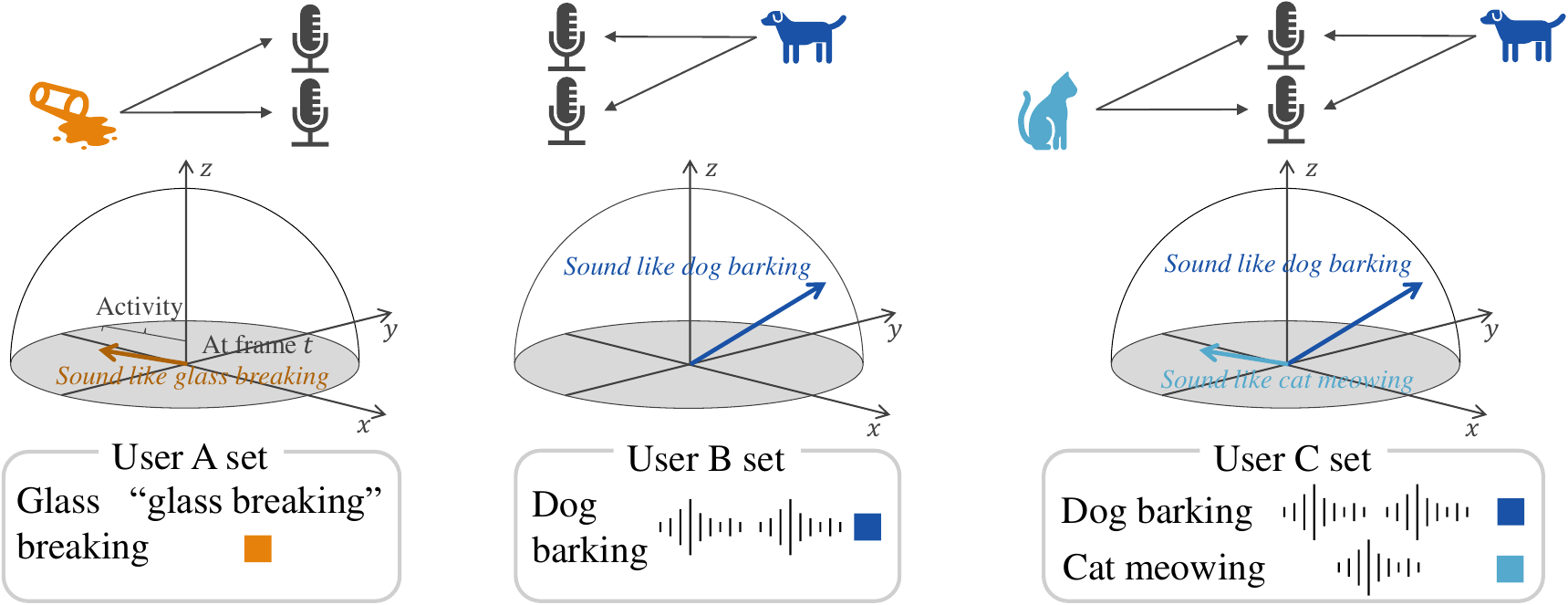}}
    \vspace{-2mm}
    \caption{
    \red{Overview of zero- and few-shot SELD system.}
    }
    \label{fig:task}
    \vspace{-5mm}
\end{figure}

However, they only output the temporal activation and DOA of preset classes trained before inference.
This is problematic because users generally prefer their own set of target sound event classes.
To customize target classes after training, we tackle zero- and few-shot SELD tasks, in which we set new classes with zero- and few-shot samples (Fig.~\ref{fig:task}).
In this context, a zero-shot sample means a text sample of sound events, e.g., “glass breaking,” and few-shot samples mean a few audio samples of sound events.

\begin{figure}[t]
    \centering
    \centerline{\includegraphics[width=0.89\linewidth]{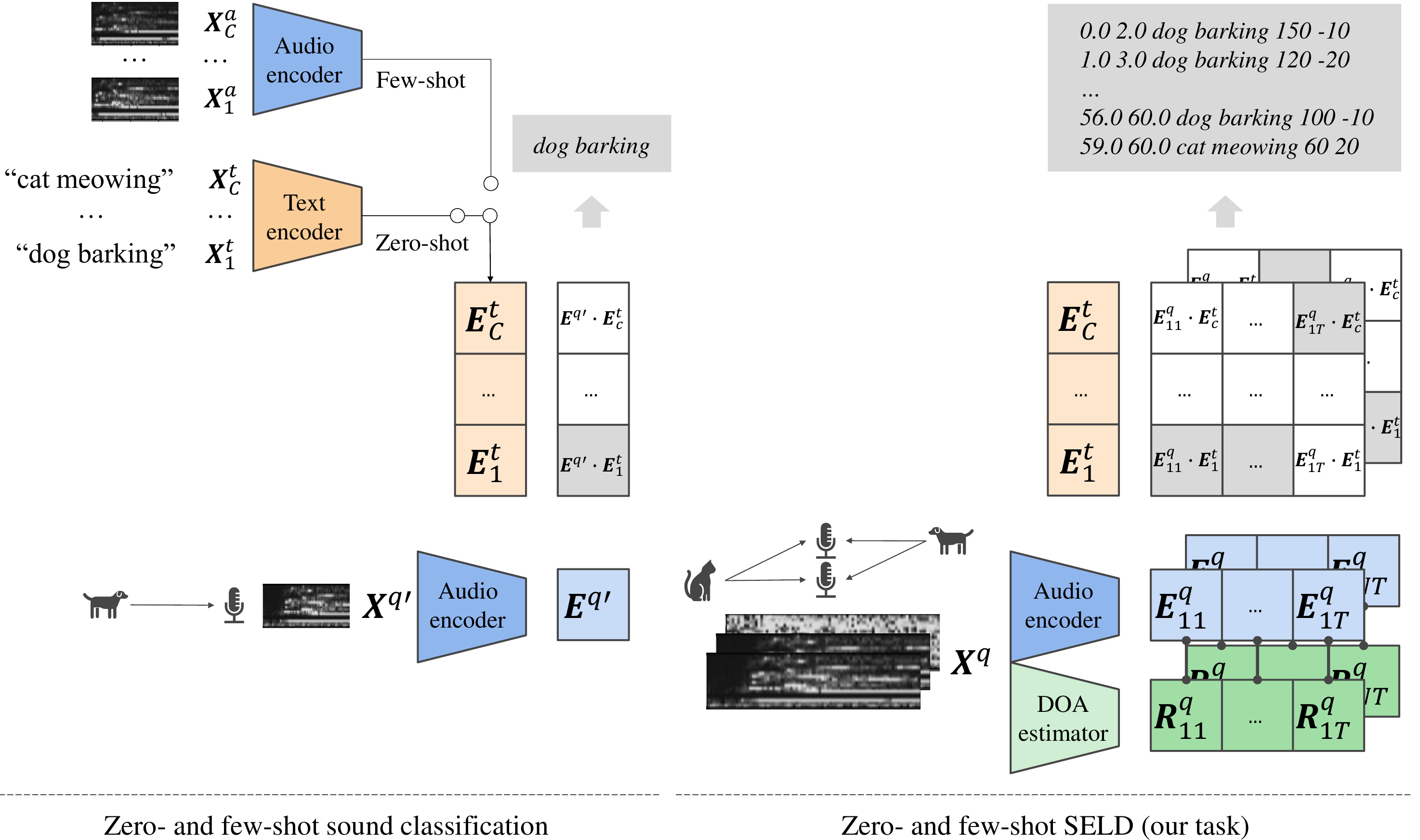}}
    \vspace{-2mm}
    \caption{
    \red{Zero- and few-shot sound classification and SELD tasks.}
    }
    \label{fig:zerofewtasks}
    \vspace{-5mm}
\end{figure}

\red{Contrastive language-audio pretraining~(CLAP) allows for zero-shot tasks~\cite{elizalde2023clap,wu2023large,kushwaha2023multimodal}, similar to contrastive language-image pretraining~(CLIP)~\cite{radford2021learning,gu2022open}.}
CLAP learns to connect language and audio using two encoders and contrastive learning to bring audio and text descriptions into a joint multi-modal space.
Zero-shot classification is solved by computing the cosine similarity between the CLAP embeddings of an audio query and text support samples~\cite{elizalde2023clap,wu2023large,kushwaha2023multimodal}.
When we have a few audio samples of target classes, we can tackle few-shot audio tasks.
In addition to the few-shot classification task~\cite{chou2019learning}, several works have tackled the few-shot sound event detection~(SED) task~\cite{wang2020few,shimada2020metric}.
A few-shot SED system takes an audio query sequence and needs to estimate sections without target classes, i.e., background noise sections~\cite{shimada2020metric}.
While the CLAP embeddings allow us to tackle zero- and few-shot classification and SED tasks, zero- and few-shot SELD tasks have other requirements: specifically, they must output an event's embedding and its corresponding DOA in both single source and overlapping cases, as shown in Fig.~\ref{fig:zerofewtasks}.

In this paper, to solve the assignment problem in overlapping cases, we propose an embed-ACCDOA model, which is trained to output track-wise CLAP embedding and the corresponding ACCDOA vector.
The embed-ACCDOA model uses a similar network architecture to the track-wise SELD method EINV2~\cite{cao2021improved}.
Before inference, we obtain support embeddings from zero- or few-shot samples of target classes.
In the inference, if activity from the estimated ACCDOA vector is larger than a threshold, the system outputs DOA from the ACCDOA vector and a class whose support embedding is the nearest to the estimated embedding.
We also propose combining the embed-ACCDOA model and the CLAP audio encoder to utilize a CLAP embedding itself in single-source cases.
To investigate the proposed methods in zero- and few-shot SELD tasks, we first train the embed-ACCDOA model with a synthetic dataset, and then evaluate the system with two datasets: Sony-TAU Realistic Spatial Soundscapes 2023~(STARSS23)~\cite{politis2022starss22,shimada2023starss23} and TAU-NIGENS Spatial Sound Events 2021~(TNSSE21)~\cite{politis2021dataset}.
We prepare a straightforward combination of the CLAP audio encoder and a DOA estimation~(DOAE) model for comparison.
We also compare with the official baseline SELD systems trained with full training datasets for reference.

\section{Zero- and Few-shot SELD tasks}
\label{sec:task}

We first explain zero-shot sound classification tasks using language-audio models such as CLAP~\cite{elizalde2023clap,wu2023large,kushwaha2023multimodal}.
The left side of Fig.~\ref{fig:zerofewtasks} shows a zero-shot sound classification task.
We define a zero-shot support sample of class $c$ as a text data $\Vec{X}^{t}_{c}$, e.g., “\texttt{class-name}.”
Let an audio spectrogram $\Vec{X}^{q'} \in {\mathbb{R}}^{F' \times T'}$ be a query sample.
$F'$ and $T'$ indicate the numbers of frequency bins and time frames, respectively.
The $D$-dimensional embedding of text support $\Vec{E}^{t}_{c} \in {\mathbb{R}}^{D}$ and one of audio query $\Vec{E}^{q'} \in {\mathbb{R}}^{D}$ are respectively obtained by a CLAP text encoder $\mathcal{F}_\mathrm{CLAPtext}$ and a CLAP audio encoder $\mathcal{F}_\mathrm{CLAPaudio}$:
\begin{align}
    \Vec{E}^{t}_{c} &= \mathcal{F}_\mathrm{CLAPtext}(\Vec{X}^{t}_{c}), \\
    \Vec{E}^{q'} &= \mathcal{F}_\mathrm{CLAPaudio}(\Vec{X}^{q'}).
    \label{eq:zero_classification}
\end{align}
For $C$ classes, we construct $C$ prompt texts $\Vec{X}^{t} = \{ \Vec{X}^{t}_{c} \}_{c = 1, ..., C}$.
For a given audio $\Vec{X}^{q'}$, we determine the best match $\Vec{X}^{t}_{c}$ among $\Vec{X}^{t}$ by the cosine similarity function over their embeddings $\Vec{E}^{q'}$ and $\Vec{E}^{t}_{c}$.

We keep the same support embeddings $\Vec{E}^{t}_{c}$ for the zero-shot SELD tasks.
Unlike zero-shot sound classification tasks, zero-shot SELD tasks need to output per time frame, to consider overlapping sound events, and to associate embeddings and DOAs.
The right side of Fig.~\ref{fig:zerofewtasks} depicts a zero-shot SELD task, given a multichannel audio query sequence $\Vec{X}^{q} \in {\mathbb{R}}^{M \times F \times T}$, where $M$, $F$, and $T$ indicate the numbers of feature channels, frequency bins, and time frames, respectively.
To solve the requirements for time frames and overlapping sound events, the audio query embeddings should have dimensions of time and track, i.e., $\Vec{E}^{q} \in {\mathbb{R}}^{D \times N \times T}$, where $N$ indicates the numbers of output tracks.
Also, audio query embeddings $\Vec{E}^{q}$ need to be associated with Cartesian DOA vectors $\Vec{R}^{q} \in {\mathbb{R}}^{3 \times N \times T}$.
\red{For each audio query embedding $\Vec{E}^{q}_{nt}$, we determine the best match $\Vec{E}^{t}_{c}$ among $\Vec{E}^{t} = \{ \Vec{E}^{t}_{c} \}_{c = 1, ..., C}$ by the cosine similarity function.}

When we use a $K$-shot audio support set $\Vec{X}^{a}_{c} = \{ \Vec{X}^{a}_{ck} \in {\mathbb{R}}^{F' \times T'} \}_{k = 1, ..., K}$ instead of a zero-shot text sample for a class ${c}$, we replace the text embedding $\Vec{E}^{t}_{c}$ with an audio embedding $\Vec{E}^{a}_{c} \in {\mathbb{R}}^{D}$:
\begin{align}
    \Vec{E}^{a}_{c} &=  \frac{1}{K} \sum_{\Vec{X}^{a}_{ck} \in \Vec{X}^{a}_{c}} \mathcal{F}_\mathrm{CLAPaudio}(\Vec{X}^{a}_{ck}),
    \label{eq:few_classification}
\end{align}
where we use an average of the embeddings called a prototype~\cite{snell2017prototypical}.

\section{Method}
\label{sec:method}

\subsection{Embed-ACCDOA model}
\label{ssec:embed_accdoa}

To achieve zero-shot SELD tasks, a CLAP model can output embeddings in single source cases.
However, CLAP models are not designed to output each embedding of sound events in overlapping cases.
To obtain embeddings and the corresponding DOAs in overlapping cases, we propose an embed-ACCDOA model, which outputs an embedding and an ACCDOA vector in each track.
The model is trained to output an oracle CLAP embedding and its corresponding ACCDOA vector in a track, given a multichannel spectrogram $\Vec{X} \in {\mathbb{R}}^{M \times F \times T}$.
The embed-ACCDOA format is formulated by embeddings, $\Vec{E} \in {\mathbb{R}}^{D \times N \times T}$, and ACCDOA vectors, $\Vec{P} \in {\mathbb{R}}^{3 \times N \times T}$.

Each ACCDOA vector of a track is represented by three nodes corresponding to the sound event location in the $x$, $y$, and $z$ axes~\cite{shimada2021accdoa}.
Let $\Vec{a} \in {\mathbb{R}}^{N \times T}$ be activities, whose reference value is ${a}_{nt}^{*} \in \{0, 1\}$, i.e., it is 1 when the event is active and 0 when inactive.
Also, let $\Vec{R} \in {\mathbb{R}}^{3 \times N \times T}$ be Cartesian DOAs, where the length of each Cartesian DOA is 1, i.e., $||\Vec{R}_{nt}|| = 1$ when a track $n$ is active.
$||\cdot||$~is the L2 norm.
An ACCDOA vector is formulated as follows~\cite{shimada2021accdoa}:
\begin{align}
    \Vec{P}_{nt} =
        {a}_{nt} \Vec{R}_{nt}.
    \label{eq:accdoa}
\end{align}
An activity and a Cartesian DOA vector are obtained from the ACCDOA vector~\cite{shimada2021accdoa}:
\begin{align}
    {a}_{nt} &= ||\Vec{P}_{nt}||, \\
    \Vec{R}_{nt} &= \frac{\Vec{P}_{nt}}{||\Vec{P}_{nt}||}.
    \label{eq:activity_DOA}
\end{align}

In training, we use synthetic mixture with ${J}$ clean directional sound events $\{ \Vec{X}_{j} \in {\mathbb{R}}^{M \times F \times T} \}_{j = 1, ..., J}$:
\begin{align}
    \Vec{X} = \sum_{j}^{J} \Vec{X}_{j} + \Vec{N},
    \label{eq:mixture}
\end{align}
where $\Vec{N} \in {\mathbb{R}}^{M \times F \times T}$ is ambient noise.
To obtain an oracle embedding $\Vec{E}^{*}_{jt}$, we use CLAP audio embeddings from clean events:
\begin{align}
    \Vec{E}^{*}_{jt} = \mathcal{F}_\mathrm{CLAPaudio}(\Vec{X}_{j}^{(1)}),
    \label{eq:clap}
\end{align}
where we utilize the entire length event of the first channel $\Vec{X}_{j}^{(1)}$ since one frame of the spectrogram is too short to obtain an accurate CLAP embedding.
If the number of events is less than the number of tracks, we set a zero vector as an oracle embedding.

\begin{figure}[t]
    \centering
    \centerline{\includegraphics[width=0.83\linewidth]{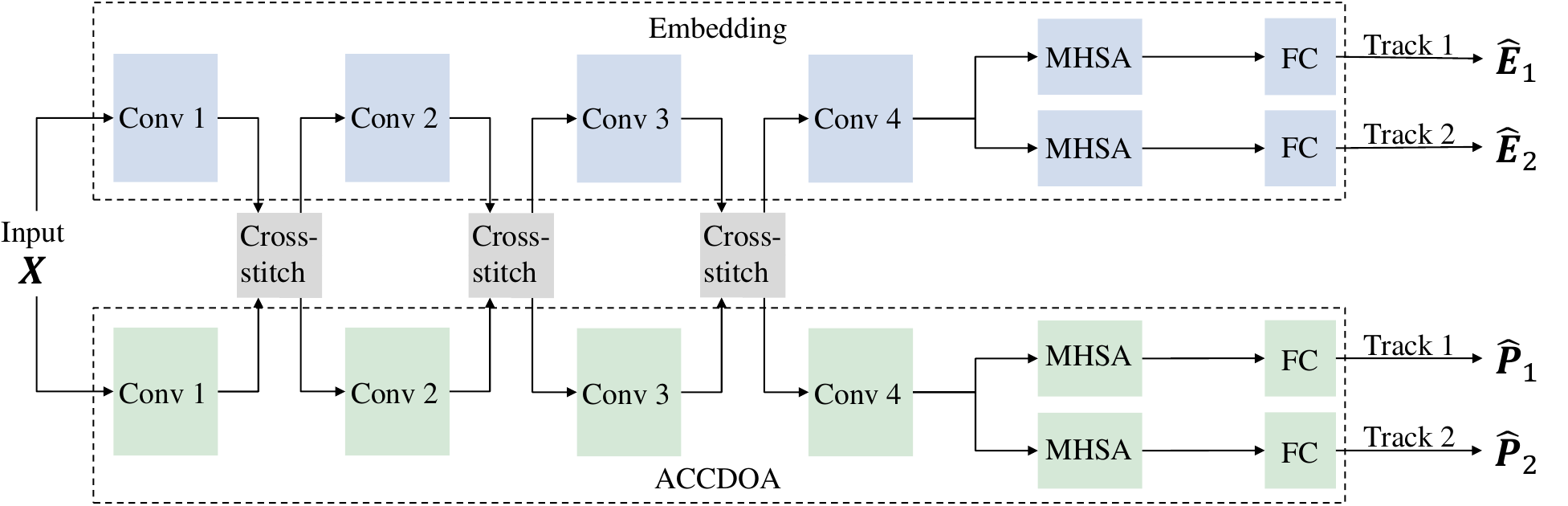}}
    \vspace{-2mm}
    \caption{
    \red{Overview of a 2-track embed-ACCDOA model.}
    }
    \label{fig:embed_accdoa}
    \vspace{-5mm}
\end{figure}

To output embeddings and ACCDOA vectors in a track-wise manner, the embed-ACCDOA model uses a similar network architecture to EINV2~\cite{cao2021improved}, a well-known track-wise model.
Our architecture has two branches: an embedding branch and an ACCDOA branch.
Each branch consists of convolution blocks, multi-head self-attention (MHSA) blocks, and fully connected layers.
Also, the convolution blocks use cross-stitch \red{units}~\cite{cao2021improved,misra2016cross} to share parameters between the two branches.
The architecture is depicted in Fig.~\ref{fig:embed_accdoa}.

Similarly to other track-wise approaches, the embed-ACCDOA model also suffers from the track permutation problem.
To overcome this issue, we adopt permutation-invariant training~(PIT) for the training process.
The frame-level PIT~\cite{cao2021improved} is used in this study.
Assume all possible permutations constitute a permutation set $\mathrm{Perm}$.
$\alpha \in \mathrm{Perm}(t)$ is one possible frame-level permutation at frame $t$.
A PIT loss for the embed-ACCDOA format can be written as follows:
\begin{align}
    \mathcal{L}^\mathrm{PIT} &= \frac{1}{T} \sum_{t}^{T} \min_{\alpha \in \mathrm{Perm}(t)} {l}^{\mathrm{EA}}_{\alpha, t}, 
    \label{eq:loss_frame_pit} \\
    {l}^\mathrm{EA}_{\alpha, t} &= \frac{1}{N} \sum_{n}^{N} {\beta}_{E} {l}^\mathrm{E}_{\alpha, nt} + {\beta}_{A} {l}^\mathrm{A}_{\alpha, nt},
    \label{eq:loss_frame_embed_accdoa} \\
    {l}^\mathrm{E}_{\alpha, nt} &= \mathrm{CosineSimilarity}(\Vec{E}_{\alpha, nt}^{*}, \hat{\Vec{E}}_{nt}),
    \label{eq:loss_frame_embed} \\
    {l}^\mathrm{A}_{\alpha, nt} &= \mathrm{MSE}(\Vec{P}_{\alpha, nt}^{*}, \hat{\Vec{P}}_{nt}),
    \label{eq:loss_frame_accdoa}
\end{align}
where $\Vec{E}_{\alpha, nt}^{*}$ and $\Vec{P}_{\alpha, nt}^{*}$ are respectively an oracle embedding and ACCDOA vector of a permutation $\alpha$, and $\hat{\Vec{E}}_{nt}$ and $\hat{\Vec{P}}_{nt}$ are respectively a predicted embedding and ACCDOA vector, at track $n$ and frame $t$.
${\beta}_{E}$ and ${\beta}_{A}$ are loss coefficients for embedding and ACCDOA, respectively.
We use cosine similarity as a loss function for embeddings, as it is widely used in zero-shot tasks.
We use mean squared error~(MSE) as a loss function for the ACCDOA vectors~\cite{shimada2021accdoa}.

In inference, we prepare each support embedding $\Vec{E}^{s}_{c} \in \{ \Vec{E}^{t}_{c}, \Vec{E}^{a}_{c} \}$ as described in Section~\ref{sec:task}.
Given a multichannel query sequence $\Vec{X}^{q}$, the embed-ACCDOA model estimates embedding $\hat{\Vec{E}}$ and the corresponding ACCDOA $\hat{\Vec{P}}$.
Class at track $n$ and frame $t$, $\hat{c}_{nt}$, is obtained from the embedding as follows:
\begin{align}
    \hat{c}_{nt} = \Argmax_{c \in C} \mathrm{CosineSimilarity}(\hat{\Vec{E}}_{nt}, \Vec{E}^{s}_{c}).
    \label{eq:class}
\end{align}
To obtain the final outputs of class and DOA at frame $t$, $(\hat{c}_{t}, \hat{\Vec{R}}_{t})^{l}$, where $l \in \{0, 1, ..., n\}$, we use two thresholds: ${\sigma}_{a}$ for the track with the highest activity, and ${\sigma}_{b} (> {\sigma}_{a})$ for other tracks.
\red{
\begin{align}
    \mathrm{Threshold}_{nt} = 
    \begin{cases}
        {\sigma}_{a} & \text{if ${n} = \Argmax_{n' \in N} {a}_{n't}$}, \\
        {\sigma}_{b} & \text{if ${n} \neq \Argmax_{n' \in N} {a}_{n't}$}.
    \end{cases}
    \label{eq:threshold}
\end{align}
}
Since single source cases are easier to estimate than overlapping cases, a lower ${\sigma}_{a}$ can increase true positives while a higher ${\sigma}_{b}$ can prevent false positives.

We also incorporate a support embedding for background noise $\Vec{E}^{s}_\mathrm{noise}$ to decrease false positives.
If the support embedding for noise is more similar to an estimated embedding than the target classes, we set no event at the frame.
In a zero-shot setting, we obtain the support embedding with the text data of “silent.”
We take a few audio samples without target classes in a few-shot setting.

\subsection{Combination of CLAP and Embed-ACCDOA}
\label{ssec:clap_embed_accdoa}

To improve the zero-shot SELD performance in single source cases, we combine the embed-ACCDOA model and the CLAP audio encoder in inference.
When there is only one source after the thresholding, we calculate the cosine similarity between the support embeddings and the embedding from the CLAP encoder $\hat{\Vec{E}}_{t,\mathrm{CLAP}}$ \red{instead of the embedding predicted by embed-ACCDOA}.
We finally use the class $\hat{c}_{t,\mathrm{CLAP}}$, which is calculated as
\begin{align}
    \hat{\Vec{E}}_{t,\mathrm{CLAP}} & = \mathcal{F}_\mathrm{CLAPaudio}(\Vec{X}^{q,(1)}), \\
    \hat{c}_{t,\mathrm{CLAP}} & = \Argmax_{c \in C} \mathrm{CosineSimilarity}(\hat{\Vec{E}}_{t,\mathrm{CLAP}}, \Vec{E}^{s}_{c}),
    \label{eq:class_clap}
\end{align}
where we utilize the entire length signal of the first channel $\Vec{X}^{q,(1)}$, following the training phase.

\section{Experimental evaluations}
\label{sec:exp}

\begin{table*}[t]
    \centering
    \caption{SELD performance of zero- and few-shot methods evaluated for the test split of the STARSS23 development set.}
    \scalebox{0.70}{
        \begin{tabular}{l|c|ccccc}
        \toprule
        Method & \# of shots & $\rm{{ER}_{20^{\circ}}}$ & $\rm{{F}_{20^{\circ}}}$ & $\rm{{LE}_{CD}}$ & $\rm{{LR}_{CD}}$ & $\rm{\mathcal{E}_{SELD}}$ \\
        \midrule
        Combination of CLAP and DOAE                    & Zero   & 0.860 & 11.2 & 38.4 & 40.8 & 0.638 \\
                                                        & Few 10 & 0.837 & 14.3 & 36.2 & 46.5 & 0.607 \\
        \midrule
        Embed-ACCDOA (proposed)                         & Zero   & 0.835 & 15.8 & 55.2 & 29.9 & 0.671 \\
                                                        & Few 10 & 0.777 & 19.3 & 27.0 & 34.1 & 0.598 \\
        \midrule
        Combination of CLAP and Embed-ACCDOA (proposed) & Zero   & 0.773 & 18.7 & 51.9 & 36.1 & 0.628 \\
                                                        & Few 10 & 0.756 & 19.2 & 35.0 & 40.2 & 0.589 \\
        \midrule
        Official baseline trained with full training dataset  & (Full) & (0.594) & (29.4) & (23.4) & (49.8) & (0.483) \\
        \bottomrule
        \end{tabular}
    }
    \label{tb:result_starss23}
    \vspace{-5mm}
\end{table*}

We evaluate the embed-ACCDOA methods in zero- and few-shot SELD tasks using multichannel audio data with the first-order Ambisonics (FOA) format.
We compare the proposed methods with a straightforward combination of a CLAP audio encoder and a DOAE model.
The proposed methods are also compared with the official baseline SELD systems trained with full training datasets.

\subsection{Task setups}
\label{ssec:tasks}

To set up the zero- and few-shot SELD tasks, we prepare training data using a data generator\footnote{https://github.com/danielkrause/DCASE2022-data-generator} from the TAU Spatial Room Impulse Response Database~(TAU-SRIR DB)\footnote{https://zenodo.org/record/6408611} and Freesound Dataset 50k~(FSD50K)~\cite{fonseca2021fsd50k}.
The data generator and the data of spatial room impulse response~(SRIR) and noise are used to synthesize a part of the training data for DCASE2023 Challenge Task 3.
While the synthetic data for the challenges chose FSD50K samples to match the target classes, our training data for zero- and few-shot SELD tasks used all the FSD50K training samples.
Finally, 2,250 one-minute spatial mixtures are synthesized using the measured SRIRs and noise from nine rooms and the samples from FSD50K.

As an evaluation dataset for the zero- and few-shot SELD tasks, we use the development set of STARSS23~\cite{politis2022starss22,shimada2023starss23}.
The recordings contain 13 target sound event classes such as footsteps and bell.
The development set of STARSS23 totals about 7 hours and 22 minutes, of which 168 clips are recorded with 57 participants in 16 rooms.
The development set is further split into \texttt{dev-set-train} (90 clips) and \texttt{dev-set-test} (78 clips).
$K$-shot audio samples of the 13 target classes are extracted from \texttt{dev-set-train} in the few-shot setting, while the zero-shot setting does not use audio samples.
We use \texttt{dev-set-test} as query sequences in the evaluation.

The proposed methods can use other sets of target classes without re-training.
To check the performance in another dataset, we prepare the development set of TNSSE21~\cite{politis2021dataset} as an additional evaluation dataset.
The data are synthesized by adding sound event samples convolved with SRIR to spatial ambient noise.
The SRIRs and ambient noise recordings are collected at 15 different indoor locations.
The sound event samples consist of 12 event classes, such as crying baby and barking dog.
The dataset contains 600 one-minute sound scene recordings: 400 for training, 100 for validation, and 100 for testing.
$K$-shot audio samples are extracted from the training split in the few-shot setting.
We omit the validation split and use the test split as query sequences.

Following the setup, five metrics are used for the evaluation~\cite{mesaros2019joint}.
The first two metrics are the location-dependent error rate ${ER}_{20^{\circ}}$ and F-score ${F}_{20^{\circ}}$, where predictions are considered true positives only when the distance from the reference is less than $20^{\circ}$.
The next is the localization error ${LE}_{CD}$, which expresses the average angular distance between the same class's predictions and references.
The fourth is a simple localization recall metric ${LR}_{CD}$, which tells the true positive rate of how many of these localization estimates are detected in a class out of the total number of class instances.
We also adopt an aggregated SELD error, $\rm{\mathcal{E}_{SELD}}$, which is defined as
\begin{align}
    \rm{\mathcal{E}_{SELD}} = \frac{\rm{{ER}_{20^{\circ}}} + ( 1 - \rm{{F}_{20^{\circ}}} ) + \frac{\rm{{LE}_{CD}}}{\red{180^{\circ}}} + ( 1 - \rm{{LR}_{CD}} )}{4}.
    \label{eq:seld_error}
\end{align}

\subsection{Experimental settings}
\label{ssec:setting}

The embed-ACCDOA model uses the EINV2 network architecture~\cite{cao2021improved} with slight modification.
The difference between the original and the one used here is the output size of the final fully connected layer in the embedding branch, i.e., from the number of classes to the embedding size.
The embedding size is 512, following a previous CLAP implementation~\cite{wu2023large}.
The number of tracks in the embed-ACCDOA format is fixed at 3.
The loss coefficients for embedding and ACCDOA are set to 0.6 and 0.4, respectively.
We set the threshold for the track with the highest activity to 0.2.
We also set the threshold for the other tracks to 0.8.

We compare the embed-ACCDOA methods with a straightforward combination of a CLAP audio encoder~\cite{wu2023large} and a DOAE model with a one-track one-class ACCDOA format.
Since the DOAE model outputs only one ACCDOA vector per time frame, the network architecture of the model is set equivalent to the ACCDOA branch of the one-track embed-ACCDOA model.
The training data is the same.
The loss function is MSE between the oracle and the estimated ACCDOA vectors.
In inference, we simply assign the ACCDOA vector output to the class output from the CLAP audio encoder at each frame.
We set the threshold to 0.2.

Other configurations mostly follow the multi-ACCDOA paper~\cite{shimada2022multi} in all methods.
Multichannel amplitude spectrograms and inter-channel phase differences~(IPDs) are used as features.
Two data augmentation methods are applied: equalized mixture data augmentation~(EMDA)~\cite{takahashi2017aenet} and rotation in FOA~\cite{mazzon2019first}.
The sampling frequency is set to 24 kHz.
The short-term Fourier transform~(STFT) is applied with a 20-ms frame length and a 10-ms frame hop.
Input features are segmented to have a fixed length of 1.27 seconds.
The shift length is set to 1.2 seconds during inference.
We use a batch size of~32, and each training sample is generated on the fly.
We use the Adam optimizer with a weight decay of~$10^{-6}$.
We gradually increase the learning rate to 0.001 with 25,000 iterations~\cite{goyal2017accurate}.
After the warm-up, the learning rate is decreased by 10\% if the SELD error of the validation did not improve in 20,000 consecutive iterations.
We validate and save model weights every 5,000 iterations up to 200,000.
Finally, we apply stochastic weight averaging~(SWA)~\cite{izmailov2018averaging} to the last ten models.

We also run the official baseline systems for reference~\cite{politis2021dataset}.
Note that the baseline systems are trained with full training datasets, while our methods take only zero- or few-shot samples of the target classes.

\subsection{Experimental results}
\label{ssec:res}

Table~\ref{tb:result_starss23} summarizes the performance of the zero- and few-shot methods in the development set of STARSS23.
The embed-ACCDOA model shows a similar SELD error to the combination of the CLAP audio encoder and the DOAE model.
While the combination of DOAE and CLAP performs better in localization recall, the embed-ACCDOA model performs better in the location-dependent error rate and F-score.
The proposed combination of the CLAP audio encoder and the embed-ACCDOA model achieves the best SELD error among the zero- and few-shot methods.
The proposed combination tackles overlapping cases while borrowing the detection performance of the CLAP audio encoder in single-source cases.
While gaps exist between the official baseline system trained with complete train data and the zero- and few-shot systems, the proposed methods show promising results without re-training on the target classes.

\begin{figure}[t]
    \centering
    \centerline{\includegraphics[width=0.97\linewidth]{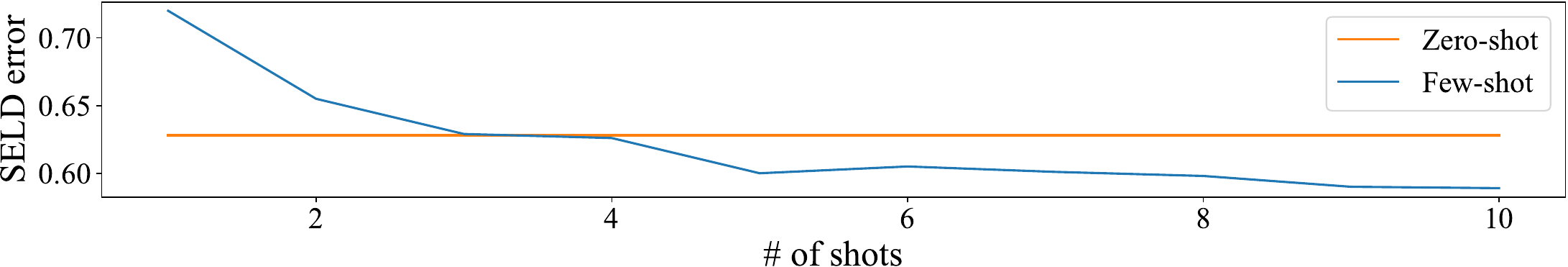}}
    \vspace{-2mm}
    \caption{
    \red{SELD performance of the combination of the CLAP audio encoder and embed-ACCDOA model for STARSS23 with different numbers of shots.}
    }
    \label{fig:graph_shots}
    \vspace{-5mm}
\end{figure}

\red{
Fig.~\ref{fig:graph_shots} shows the performance of the combination of CLAP and embed-ACCDOA with different numbers of shots.
When the number increases, the method improves performance.
The zero-shot performance is better than the 1-shot and comparable to the 3-shot.
}

Table~\ref{tb:result_tnsse21} lists the performance of the zero- and few-shot SELD methods in the development set of TNSSE21.
We use the same model as the experiments on STARSS23 to check the capability of adapting to another dataset with zero- and few-shot samples.
The proposed method without re-training achieves comparable results to the official baseline system trained with complete train data.

\begin{table}[t]
    \centering
    \caption{SELD performance of zero- and few-shot methods evaluated for the test split of the TNSSE21 development set.}
    \scalebox{0.70}{
        \begin{tabular}{l|c|ccccc}
        \toprule
        Method & \# of shots & $\rm{{ER}_{20^{\circ}}}$ & $\rm{{F}_{20^{\circ}}}$ & $\rm{{LE}_{CD}}$ & $\rm{{LR}_{CD}}$ & $\rm{\mathcal{E}_{SELD}}$ \\
        \midrule
        CLAP and Embed-ACCDOA           & Zero   & 1.008 & 25.8 & 26.2 & 46.9 & 0.607 \\
                                        & Few 10 & 1.009 & 24.9 & 31.0 & 51.5 & 0.604 \\
        \midrule
        Official baseline               & (Full) & (0.706) & (26.0) & (37.0) & (40.4) & (0.562) \\
        \bottomrule
        \end{tabular}
    }
    \label{tb:result_tnsse21}
    \vspace{-5mm}
\end{table}

\section{Conclusion}
\label{sec:conclusion}

We investigate zero- and few-shot sound event localization and detection~(SELD) tasks, which enable us to customize the target classes of SELD systems with only a text sample or a few audio samples.
While zero-shot sound classification tasks are achieved by embeddings from contrastive language-audio pretraining~(CLAP) models, zero-shot SELD tasks require the assignment of an activity and a direction-of-arrival~(DOA) to each embedding, especially in overlapping cases.
To address the assignment problem in overlapping cases, we propose an embed-ACCDOA model, which is trained to output track-wise CLAP embedding and associated activity-coupled Cartesian DOA (ACCDOA).
In our experimental evaluations on the zero- and few-shot SELD tasks, the embed-ACCDOA model shows a better location-dependent error rate and F-score than the straightforward combination of the CLAP audio encoder and the DOA estimation model.
Moreover, the proposed combination of the embed-ACCDOA model and the CLAP audio encoder with zero- or few-shot samples shows comparable performance to the official baseline system trained with complete train data in an evaluation dataset.
We will conduct comprehensive experiments in the future.

\bibliographystyle{IEEEtran}
\bibliography{refs}

\end{document}